\documentclass[aps,twocolumn,superscriptaddress]{revtex4-1}

\usepackage{graphicx,subfigure}
\makeatletter
\DeclareRobustCommand*\textsubscript[1]{%
  \@textsubscript{\selectfont#1}}
\def\@textsubscript#1{%
  {\m@th\ensuremath{_{\mbox{\fontsize\sf@size\z@#1}}}}}
\makeatother
\usepackage{dcolumn}
\usepackage{bm}

\begin{document}

\title{Electric field induced birefringence in non-aqueous dispersions of mineral nanorods}

 \author{Alexis de la Cotte}
 \author{Pascal Merzeau}
 \affiliation{Centre de Recherche Paul-Pascal, CNRS - Universit\'{e} de Bordeaux, 115 Avenue Schweitzer, 33600 Pessac, France}
 \author{Jong Wook Kim}
 \affiliation{Laboratoire de Physique de la Mati\`{e}re Condens\'{e}e, CNRS - Ecole Polytechnique, 91128 Palaiseau, France }
 \author{Khalid Lahlil}
 \author{Jean-Pierre Boilot}
 \author{Thierry Gacoin}
 \affiliation{Laboratoire de Physique de la Mati\`{e}re Condens\'{e}e, CNRS - Ecole Polytechnique, 91128 Palaiseau, France }
 \author{Eric Grelet}
 \email{grelet@crpp-bordeaux.cnrs.fr}
 \affiliation{Centre de Recherche Paul-Pascal, CNRS - Universit\'{e} de Bordeaux, 115 Avenue Schweitzer, 33600 Pessac, France}

\date{\today}

\begin{abstract}
Lanthanum phosphate (LaPO\textsubscript{4}) nanorods dispersed in the non-aqueous solvent of ethylene glycol form a system exhibiting large intrinsic birefringence, high colloidal stability and the ability to self-organize into liquid crystalline phases.
In order to probe the electro-optical response of these rod dispersions we study here the electric-field-induced birefringence, also called Kerr effect, for a concentrated isotropic liquid state with an in-plane a.c. sinusoidal electric field, in conditions of directly applied (electrodes in contact with the sample) or externally applied (electrodes outside the sample cell) fields. Performing an analysis of the electric polarizability of our rod-like particles in the framework of Maxwell-Wagner-O'Konski theory, we account quantitatively for the coupling between the induced steady-state birefringence and the electric field as a function of the voltage frequency for both sample geometries. The switching time of this non-aqueous transparent system has been measured, and combined with its high Kerr coefficients and its features of optically isotropic ``off-state" and athermal phase behavior, this represents a promising proof-of-concept for the integration of anisotropic nanoparticle suspensions into a new generation of electro-optical devices.

\end{abstract}

\pacs{}
%\keywords{}

\maketitle

\section{Introduction}

The response of anisotropic fluids to external electric fields has been widely studied and has led specifically to a major technological advance, the Liquid Crystal Displays (LCD) \cite{LCD,LCDReview}. When applied to a liquid crystalline sample, the external electric field induces usually a reorientation of the mesogens along or perpendicular to the field. This electro-optical effect has been observed on a wide spectrum of materials, ranging from thermotropic molecular compounds (exhibiting a temperature dependent phase behavior) used in LCDs \cite{YanAPL2010,Yan,Ghana,KVLe,Hisakado,Nordendorf,LCD,LCDReview,Nordendorf2} to lyotropic dispersions of anisometric colloidal nanoparticles (whose phase behavior is concentration dependent) \cite{Leferink}, such as filamentous viruses \cite{DhontEPJ2010,DhontEPJ2011}, pigment particles \cite{Eremin,Eremin2}, silica rods \cite{Kuijk,Liu} or graphene oxide sheets \cite{Shen,Hong}.\\

In the case of dilute suspensions into the isotropic liquid state, applying an electric field results in an induced nematic-like organization with the emergence of an associated birefringence \cite{Dozov2011,Antonova2012,Paineau}. %Unlike thermotropic molecular systems, colloidal dispersions are usually stabilized in solution by surface charge repulsion. 
Beyond the intrinsic properties of the individual particles such as the presence of a permanent dipole or a core polarization induced by the field, several kinds of phenomenon are likely to arise to explain the coupling of the particles with the electric field \cite{Bellini99,DhontEPJ2010,DhontEPJ2011,Stoylov,Dozov2011,Antonova2012}. The first one is due to the dielectric mismatch between the particle and the solvent, creating an electrical dipole by the accumulation of dipolar bound charges at the particle/solvent interface. Similarly, when mobile charges are present in the particles, their accumulation at the interface leads to a contrast of conductivities with the surrounding electrolyte. Both mechanisms are known as the so-called Maxwell-Wagner polarizability, and are dependent on the electric field frequency. Working with charged particles usually enables the colloidal stability of their dispersions.  This also induces the distortion of the electrical double layer by the external field (as well as other related phenomenon such as the electro-osmotic flow), as first described by O'Konski \cite{O'Konski}, who shows that the polarization of the ionic cloud is often the dominating mechanism.
%To prevent the electrophoretic motion of micro- or macro-ions within the sample, alternative (a.c.) electric fields are widely used in electro-optical setups, making each of the polarization mechanisms listed above %dependent on the field frequency. At low frequencies, the polarization by the external field of both the layer of condensed ions and the electric double layer prevails, while at high frequencies, where charge motion is %greatly hindered, the dielectric mismatch between particle and solvent is the dominating mechanism \cite{DhontEPJ2010,DhontEPJ2011,Dhont2014,Stoylov}.\\ %\textbf{}

We present here a study of the electric field induced birefringence in liquid suspensions of lanthanum phosphate (LaPO\textsubscript{4}) nanorods. A key feature of our system is that while most of the studies are performed in aqueous solvent \cite{DhontEPJ2010,DhontEPJ2011,Dozov2011,Antonova2012,Paineau,Leunissen,Mantegazza,Bellini,Digiorgio,Velev,Shen,Hong}, our LaPO\textsubscript{4} rod-like particles are dispersed in ethylene glycol in which they exhibit an outstanding colloidal stability compared to water as reported in a previous paper \cite{KimAdv2012}.
In this work, we first investigate the frequency dependence of the steady-state electric birefringence in our non-aqueous suspensions. Under an in-plane alternative (a.c.) electric field and with electrodes immersed in the sample, the frequency threshold between the conductivity contrast or dielectric mismatch based polarization mechanisms is determined. %highlighting the differences in behavior between ethylene glycol- and water-based dispersions. 
We then account quantitatively for the Kerr effect and its dependence with the field frequency and with the rod volume fraction thanks to the Maxwell-Wagner-O'Konski theory. The switching time, and consequently the rotational diffusion coefficient of the LaPO\textsubscript{4} rods in ethylene glycol, is measured  in order to consider the potential use of such a system in electro-optical devices.
Due to the limitations brought by the electrophoretic motion of charges within the sample cell leading to possible electric shortcircuits, we have also designed an original cell, inspired by previous works \cite{Dozov2011,Antonova2012}, in which the electric field is applied tangentially to a thin dielectric quartz capillary wall limiting therefore its attenuation as shown by finite-element simulations. This geometry, which has the benefit to avoid any electric degradation of the sample even in steady-state regime under continuous sinusoidal field, shows similar optical retardation compared to cells with directly applied in-plane electric field. \\

\section{Materials and Setups}
\subsection{Sample preparation and characterization}

\begin{figure}
  \includegraphics{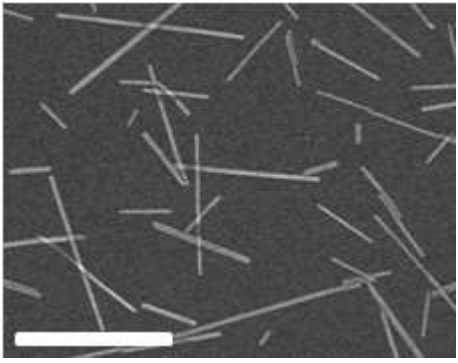}\\
  \caption{\label{Fig:SEM}Scanning electron microscopy picture of the LaPO\textsubscript{4} nanorods obtained according to ref. \cite{KimAdv2012}. The scale bar is 500~nm. }
\end{figure}

LaPO\textsubscript{4} nanorods, synthesized according to ref. \cite{KimAdv2012}, are shown in Fig. \ref{Fig:SEM}. Their main features are an average length of $L=250$~nm and a diameter of $D=11$~nm. Their polydispersity is provided through the standard deviation of their size distribution, $\sigma_L=0.63$  and $\sigma_D=0.19$, respectively \cite{KimAdv2012}. %reported in Table \ref{tab:sample}. 
The samples are prepared by diluting the rod-like particles with ethylene glycol to reach a concentration close to the binodal volume fraction ($\phi_{B}= 0.8\% \pm 0.1 \% $) corresponding to the highest concentration of the isotropic liquid phase before the transition to the nematic state. The corresponding range of ionic strength $I$ is in between 0.1 and 1~mM depending on the sample dilution. The dispersion conductivity is measured thanks to a conductivity meter (CDM210, Radiometer analytical). The LaPO\textsubscript{4} surface charge density is estimated by zeta-potential measurement (Zeta Compact, CAD Instruments) as described in the Electronic Supplementary Information (ESI).
% Two different batches studied at different time => variation in ionic strength.

% \begin{table}[b]
% \caption{\label{tab:sample} Samples of LaPO\textsubscript{4} nanorods dispersed in ethylene glycol where their initial volume fractions ($\phi_{0}$), their binodal volume fractions ($\phi_{B}$) and their core radius ($a$) and lengths ($L$) are indicated. }
% \begin{ruledtabular}
% \begin{tabular}{lcccc}
% \textrm{Sample name} & \textrm{$\phi_{0}$} & \textrm{$\phi_{B}$} & \textrm{$a$ (nm)} & \textrm{$L$ (nm)}\\
% \colrule
% Batch 1  & 3.30\%  & 0.80\% & 5.3 & 252 \\
% Batch 2 & 3.40\% & 0.91\% & 5.5 & 200 \\
% \end{tabular}
% \end{ruledtabular}
% \end{table}

\subsection{Electro-optical setups}
\begin{figure}
  \includegraphics[width=0.94\columnwidth]{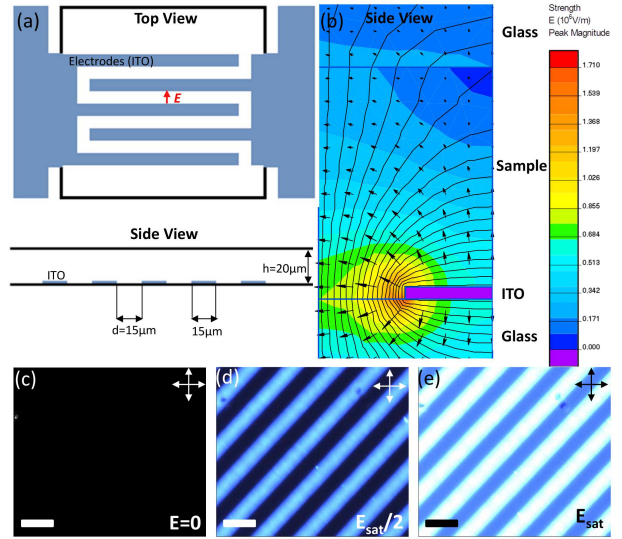}\\
  \caption{\label{Fig:IPS}(a) Schematic representation of the In-Plane Switching (IPS) cells used for probing the electro-optical effects in our colloidal dispersions. (b) Equipotential lines of the electric field ($f=100$~kHz) simulated by finite element method inside an IPS cell containing ethylene glycol. For symmetry reason, only half an electrode is represented (Side view). (c) Observation by optical microscopy between crossed polarizers of an IPS cell (top view) filled with an isotropic liquid suspension of LaPO\textsubscript{4} nanorods when no electric field is applied (off-state). (d) and (e) Progressive increase of the electric induced birefringence with the in-plane electric field (Scale bars: 30~$\mu$m).}
\end{figure}

%\subsubsection{In-Plane Switching cells}
The first electro-optical setup is a commercial In-Plane Switching (IPS) cell containing 330 line-shaped electrodes (Instec Inc). A schematic representation is given in Figure \ref{Fig:IPS}-(a). The samples are prepared by loading by capillarity the LaPO\textsubscript{4} nanorod suspensions in the cells, which are then sealed with UV glue (Epotecny) to prevent any evaporation of the ethylene glycol solvent.

\begin{figure}
 \subfigure[]{\label{A}
  \includegraphics[width=0.7\columnwidth]{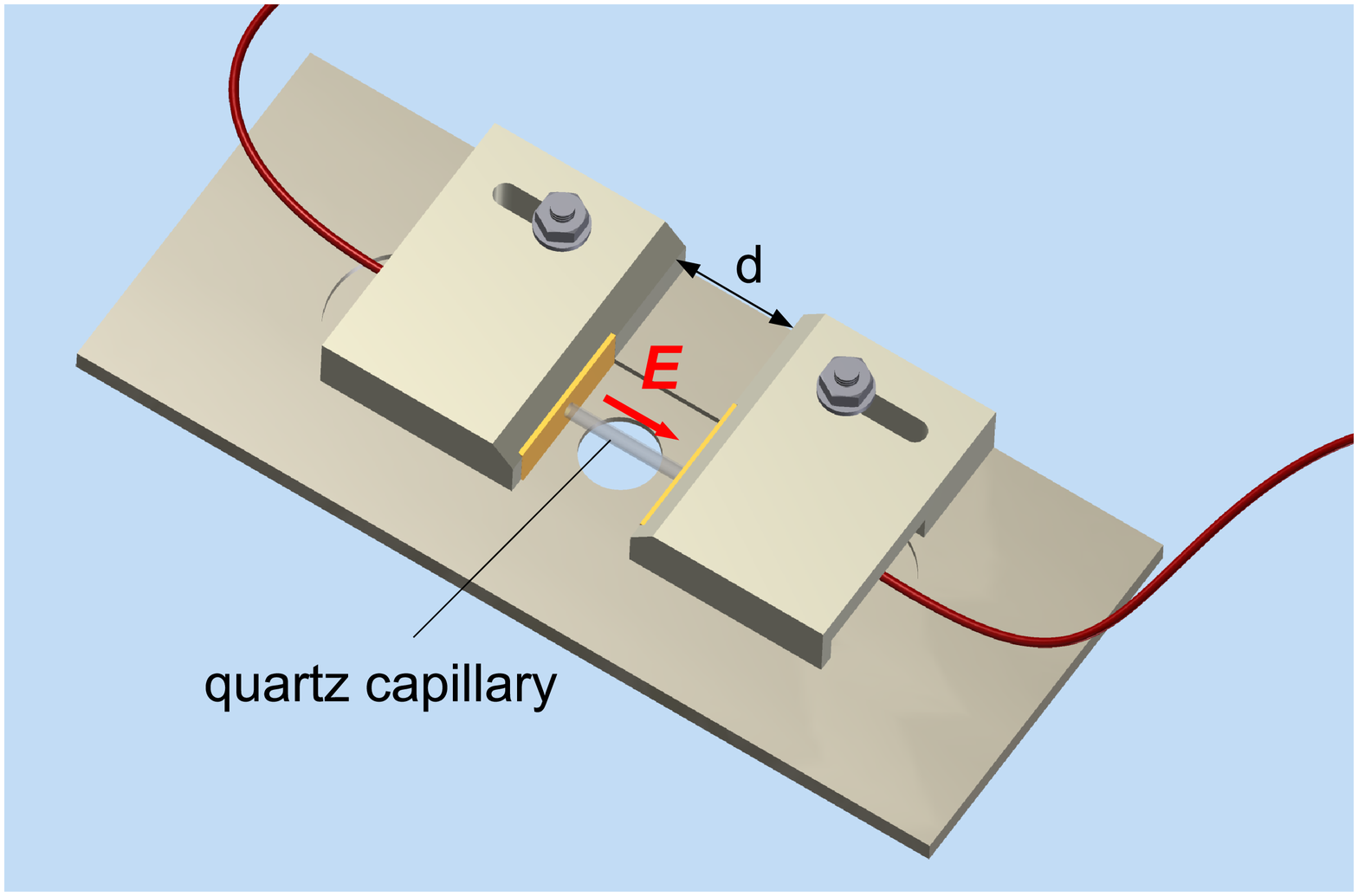}}\\
 \subfigure[]{\label{B}
  \includegraphics[width=0.95\columnwidth]{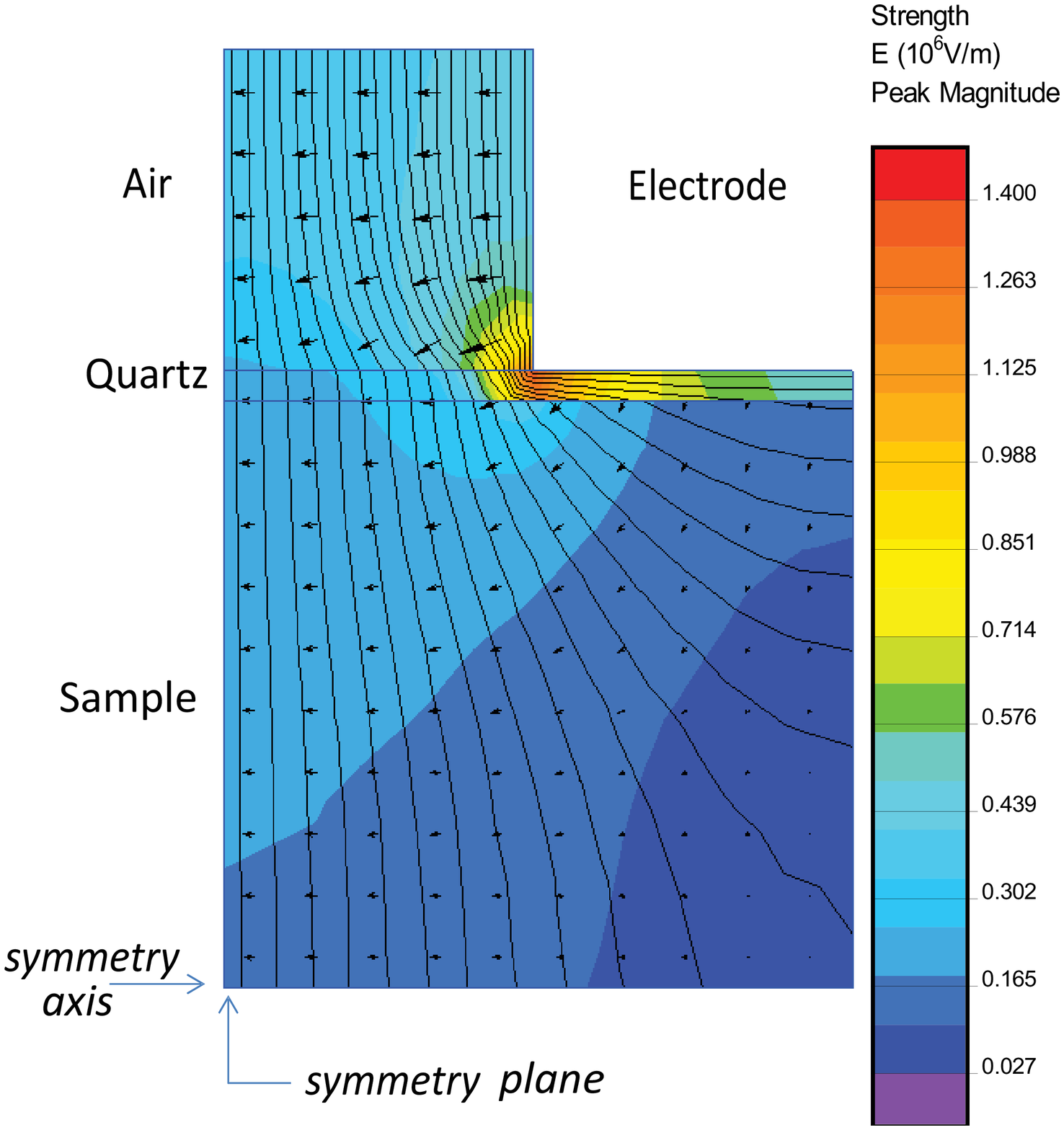}}\\
  \caption{\label{Fig:SetupCap} (a) Schematic representation of the electro-optical setup used for the study of the nanorod suspensions in capillaries. The yellow part corresponds to the electrodes made of aluminum planes (width: 20~mm, height: 6.5~mm, and thickness: 2~mm). The quartz capillary is 1~mm in diameter and the inter-electrode distance ($d$) is an experimental tunable parameter. (b) Simulations of the equipotential lines associated with the electric field ($f=100$~kHz) where only one quadrant of the setup ($d=500~\mu$m)  is shown for symmetry reason.}
\end{figure}

Electric contacts between each Indium Tin Oxide (ITO) electrode and wires are made thanks to silver lacquer. Copper tape is added on the top of each contact to ensure good mechanical resistance when the cell is manipulated. The cell is then connected to a function generator (Centrad GF265) through an electric amplifier (FLC Electronics A400DI) which provides a voltage gain of a factor 20 for frequencies up to 400 kHz.
When induced between two infinite parallel planes, the electric field ($E$) is the result of the ratio between the operating voltage ($U_{0}$) and the inter-electrode distance ($d$). Moreover, the electric field being the result of an a.c. sinusoidal signal, the calculation is performed taking the root mean square (rms) value of the applied voltage:

\begin{eqnarray}
E=
c_g
\frac{U_0}{\sqrt{2} \times d}
\label{eq:fieldeff}
\end{eqnarray}

In Eq. \ref{eq:fieldeff}, $c_g$ is a correction factor accounting for the field attenuation in the sample due to the cell geometry. For two infinite parallel electrodes,  we have $c_g = 1$. 
However, IPS cells are far from this well-defined geometry (Fig. \ref{Fig:IPS}-(a)) therefore numerical simulations of the effective electric field inside the cell have been performed. The results obtained by finite element analysis using the QuickField software are shown in Fig. \ref{Fig:IPS}-(b), providing an attenuation factor of $c_g \simeq 0.46$, which is the mean square value over the cell thickness ($h$) and measured at an equidistance of two ITO electrodes. Computer simulations show also that $c_g$ is only slightly sensitive to both the electrolyte conductivity ($K_e$) and to the field frequency ($f$). Note that $c_g$ depends on the cell thickness, and decreasing $h$ from $20~\mu  m$ as used experimentally (Fig. \ref{Fig:IPS}-(a)) to typically $5~\mu  m$ would lead to a more homogeneous electric field giving $c_g \simeq 0.86$ with, however, a lower associated optical retardation  ($\delta$). Indeed, $\delta$ is related to the induced birefringence ($\Delta n$) through the thickness of the sample by: %(Fig. \ref{Fig:IPS}-(a)):
\begin{eqnarray}
\Delta n=
\frac{\delta}{h}
\label{eq:birefringence}
\end{eqnarray} 
Experimentally, this is measured using a 0-3 $\lambda$ Berek compensator (Olympus) on cells observed by polarizing microscopy (Olympus BX51) (Fig. \ref{Fig:IPS}-(c) to (e)). In order to determine the highest electric stationary birefringence, the polarizer and analyzer are crossed at $\pm$ 45$^\circ$ with respect to the field direction.\\
%The accuracy of the birefringence measurement is therefore improved by choosing cells with high enough thickness  $20~\mu  m$ .  
%Once the rods are fully orientated with the field a saturation of the electric birefringence can be reached.
Synchronizing the on/off state of the applied voltage during observation with a CCD camera (JAI, CV-M7), the birefringence relaxation of the sample from a electric-field nematic steady-state back to its isotropic liquid phase (i.e. with no induced birefringence) has been studied. To define precisely the time at which the field is switched off, a light pulse coupled to a static relay is sent to the camera using a small LED inserted above the microscope analyzer. %The pulse duration (of about 50~ms) in the LED is chosen to affect no more than two images of the movies made at the video rate of 25 frames per second. 
The transmitted intensity  of the images recorded at the video rate of 25 frames per second, has been analyzed using an image processing software (ImageJ) allowing for the determination of the relaxation or switching time ($\tau_s$) of our colloidal suspensions.\\

Studies of the electro-optical effects in the LaPO\textsubscript{4} suspensions have been also conducted in quartz capillaries of diameter $h=1~$mm (W. M\"{u}ller, Germany) by applying the electric field through their walls (10~$\mu m $ thick) thanks to a home-made setup sketched in Fig. \ref{Fig:SetupCap}-(a). The main advantage is to prevent electric sample degradation as electrolysis, faradaic reactions or shortcircuits due to the ions in the electrolyte or to charge injection at the electrodes.
%the electrophoretic motions of counterions. 
In this specific setup, the capillary wall acting as a thin dielectric layer insulates the sample from the electrodes \cite{Dozov2011,Antonova2012}. The latter ones are made of two parallel aluminum planes with a disc-shaped hole of 1~mm of diameter in their center. Before being inserted in the setup, the capillary has been selected in order to precisely fit into the two holes, allowing for a good mechanical contact between the electrodes and the capillary wall limiting in particular any air layer in between them. %Another key feature of this setup is the tunable inter-electrode distance ($d$). 
In order to determine the field penetration inside the capillary, finite element simulations have been carried out as shown in Fig. \ref{Fig:SetupCap}-(b). An estimation of the attenuation factor ($c_g $) has therefore been obtained as a function of the applied voltage frequency ($f$) and of the inter-electrode distance ($d$), and the details of the numerical simulations can be found in the ESI. 
In particular, the correction factor, $c_g$, for our capillary geometry is almost 1 at high inter-electrode distances, providing a very homogeneous field for probing the electro-optical properties of LaPO\textsubscript{4} suspensions. However, this result depends strongly on the field frequency inside the sample, and an additional correction factor, $c_g^\prime (f)$, accounting for the accumulation of charges at the dielectric quartz wall/electrolyte interface, has to be added to the calculation of the field in Eq. \ref{eq:fieldeff}. Analytically, this factor can be approximated as \cite{Dozov2011}:

\begin{eqnarray}
c_g^\prime (f)
\approx 
\frac{f}{\sqrt{f^{2}+f_{c}^{2}}}
\label{eq:cs}
\end{eqnarray}

with $f_c$ the charge relaxation frequency of the electrolyte (See next paragraph). The frequency $f=100~$kHz at which the Kerr birefringence measurements are performed, results from a balance between both the screening of the field at low frequency due the accumulation of ions near the electrodes as well as the dielectric attenuation factor of the capillary glass wall, and the Kerr effect itself whose coupling decreases at high field frequency, as shown latter in the article. The attenuation factor is thus for this setup geometry $c_g^\prime (f=~100kHz) \simeq ~$0.43. 

\section{Experimental results and discussion}
%\subsection{Directly applied field (IPS cells)}
\subsection{Kerr effect and relaxation frequency}

\begin{figure}
  \includegraphics[width=\columnwidth]{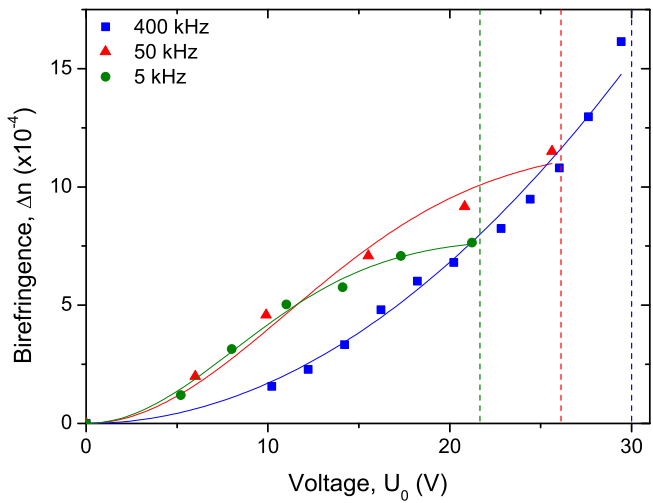}\\
  \caption{\label{Fig:Freq} Kerr induced birefringence for a sample at $\phi$=0.66\% close to the binodal volume fraction %(batch 2) 
measured with an alternative electric field in an IPS cell at three different frequencies. The symbols represent the experimental data and the solid lines the numerical fits according to equations \ref{eq:Kerrextended} and \ref{eq:Kerrparabolic} for data at 5, 50 and 400 kHz, respectively. The vertical dashed lines represent the voltage above which electric shortcircuits occur in the IPS cells. }
\end{figure}

As first qualitatively reported in Figure \ref{Fig:IPS}, the evolution of the field-induced birefringence as a  function of the applied voltage is displayed in Figure \ref{Fig:Freq}.  %with a LaPO\textsubscript{4} rod volume fraction of $\phi_{B}$=0.66\%. 
The measurements have been performed in IPS cells in the steady state regime and they have been shown to be fully reversible (by increasing or decreasing the applied voltage).

The induced birefringence and the effective electric field are classically related through a quadratic law, called Kerr effect. As the voltage is kept on increasing, all the nanorods get progressively their orientation driven by the field. The induced birefringence reaches a saturation, %($\Delta n_{sat}$), 
and it can then be expressed through the following equation \cite{YanAPL2010}:

\begin{eqnarray}
\Delta n=
\Delta n_{sat}\left(1-exp\left[\left(-
\frac{E}{E_{sat}}\right)^2\right]\right)
\label{eq:Kerrextended}
\end{eqnarray}
where $\Delta n_{sat}$ and $E_{sat}$ correspond to the saturation birefringence and field, respectively.
Expanding Eq. \ref{eq:Kerrextended} in the weak field range (i.e. $E\ll$ $E_{sat}$) provides the well-known Kerr law:

\begin{eqnarray}
\Delta n=
\lambda B E^{2}
\label{eq:Kerrparabolic}
\end{eqnarray}

where $\lambda$ is the light wavelength used for observations (set to 546 nm) %close to the highest sensitivity of the human eye) 
and $B$ is the Kerr coefficient quantifying the coupling between $\Delta n$ and $E$.
The comparison between the expansion of Eq. \ref{eq:Kerrextended} at low field and Eq. \ref{eq:Kerrparabolic} leads to the expression of the Kerr coefficient when saturation is observed in the sample:

\begin{eqnarray}
B=
\frac{\Delta n_{sat}}{\lambda \times E_{sat}^2}
\label{eq:CsteKerrsigm}
\end{eqnarray}

In Figure \ref{Fig:Freq} two kinds of behavior are displayed: at low frequencies (5 and 50 kHz) a saturation of the birefringence is observed (Eq. \ref{eq:Kerrextended}) while for the highest frequencies (400 kHz) no saturation is obtained for the same range of applied voltage (Eq. \ref{eq:Kerrparabolic}). Moreover, the voltage at which shortcircuits occur in the IPS cells increases with frequency. The decrease with the frequency of the charge mobility in the electrolyte makes more difficult to induce a continuous path of ionic charges between the electrodes. Therefore, no measurements have been performed at frequencies lower than 5~kHz where shortcircuits occur too quickly. This frequency dependence is also observed quantitatively for a pure (i.e. without LaPO\textsubscript{4} particles) electrolyte of ethylene glycol with, as expected, a shortcircuit voltage which increases by decreasing the conductivity (See additional measurements in ESI). 
It is worth mentioning that $\Delta n_{sat}$ exhibits an unexpected frequency dependence (Fig. \ref{Fig:Freq}) considering that $\Delta n_{sat}$ corresponds by definition to a perfect alignment of the rods (with an associated orientational order parameter $S_{sat}=1$). We speculate that such anomalous behavior could be attributed to some thermal disturbance stemming from charge injection at the electrodes affecting the electric field at saturation $E_{sat}$. \\   
As we shall see, the coupling of the Kerr birefringence with the electric field shown in Fig. \ref{Fig:Freq} can be quantitatively accounted by a frequency dependence of the electric polarizability of our rod-like colloidal particles formulated in the extended Maxwell-Wagner framework. 
Specifically, the induced dipole moment associated with the rod-like colloids arises from the {\it  complex} dielectric mismatch between the electrolyte and the particles. The %Maxwell-Wagner 
charge relaxation time, $\tau$, necessary for the formation of a field-induced dipole associated with the free charges (and therefore with conductivity) near the surfaces corresponds to the time for which ionic migration starts to contribute \cite{Shilov} and is therefore given by the duration of the screening process in the electrolyte \cite{Stoylov,Russel}:

\begin{eqnarray}
\tau_c 
= 
\frac{1}{D \kappa^{2}}
\label{eq:rlxtime}
\end{eqnarray}

where $D$ is the diffusion coefficient of the ions and $\kappa^{-1}$ the Debye screening length, defined as:

\begin{eqnarray}
\kappa^{-1}
=
\sqrt{
\frac{k_{B}T\varepsilon_{0}\varepsilon_{e}}{2e^2N_{A}I}
}
\label{eq:Debye}
\end{eqnarray}

with $\varepsilon_{0}$ the vacuum permittivity, $\varepsilon_{e}$ the dielectric constant of the electrolyte ($\varepsilon_{e}$=~37 for ethylene glycol), $e$ the elementary charge and $I$ the ionic strength.
The corresponding cut-off frequency, $f_{c}$, is:

\begin{eqnarray}
f_{c}
= 
\frac{1}{2\pi\tau_c}
=
\frac{D}{2\pi\kappa^{-2}}
\label{eq:fc}
\end{eqnarray} 

LaPO\textsubscript{4} nanorods are positively charged in ethylene glycol according to zeta-potential measurement ($\zeta$ = + 97 mV, See ESI), thanks to the excess of La$^{3+}$ ions close to the particle surface. When dialyzed from nitric acid solution to ethylene glycol, the ions maintaining electro-neutrality of the dispersion are therefore the counterions associated with the rod-like particles, i.e. $NO_{3}^{-}$ \cite{KimAdv2012}.
The ionic diffusion coefficient can be calculated according to:

\begin{eqnarray}
D=
\frac{k_{B}TN_{a}\mu}{|z| F}
\label{eq:diffusion}
\end{eqnarray}

with $k_{B}$ the Boltzmann constant, $N_{a}$ the Avogadro's number, $F$ the Faraday constant, $T$ the temperature (set to 298 K), $\mu$ the mobility of the ion and $z$ its valence. The mobility of the $NO_{3}^{-}$ ions is expressed as:

\begin{eqnarray}
\mu_{NO_{3}^{-}}
=
\frac{\Lambda_{NO_{3}^{-}}}{F}
\label{eq:mobility}
\end{eqnarray}

with $\Lambda_{NO_{3}^{-}}$ the molar conductivity of the ion in a given electrolyte.
In water the reported value $\Lambda_{NO_{3}^{-}}^{water}$ =7.1 $mS.m^{2}.mol^{-1}$ yields $\mu_{NO_{3}^{-}}^{water} = 7.4 \times 10^{-8}~m^{2}.V^{-1}.s^{-1}$. Assuming a simple hydrodynamic model \cite{VoronelPRL1998,KoneshanJPCB1998,KoneshanJACS1998} where this mobility is inversely proportional to the viscosity of the solvent, the  ionic mobility can be extrapolated in ethylene glycol:

\begin{eqnarray}
\mu^{EG}_{NO_{3}^{-}}=
\mu^{water}_{NO_{3}^{-}}
\frac{\eta_{water}}{\eta_{EG}}
\label{eq:mobility2}
\end{eqnarray}

with $\eta_{water}=0.9 \times 10^{-3}$~Pa.s  and $\eta_{EG}~=16 \times 10^{-3}$~Pa.s  the viscosities of water and ethylene glycol at 298~K, respectively \cite{handbook}.\\
Finally, the conductivity ($K_{e}$) of the electrolyte is introduced: 
\begin{eqnarray}
K_{e}
=
\sum\limits_{i} {c_{i}\Lambda_{i}}
=
2I\sum\limits_{i} {\Lambda_{i}}
\label{eq:Ke}
\end{eqnarray}

with $c_{i}$ the ionic molar concentration of the monovalent salt in solution. Assuming that $NO_{3}^{-}$ are the main mobile ions in the electrolyte and substituting Eqs. \ref{eq:diffusion},  \ref{eq:mobility} and \ref{eq:Debye} in Eq. \ref{eq:Ke} yields:

\begin{eqnarray}
K_e
= 
\varepsilon_{0}\varepsilon_{e} D \kappa ^2
\label{eq:fcInt}
\end{eqnarray}

which then coincides, according to Eq. \ref{eq:fc}, with the well-know formulation of the charge relaxation frequency:

\begin{eqnarray}
f_{c}
= 
\frac{K_{e}}{2\pi\varepsilon_{0}\varepsilon_{e}}
\label{eq:fcMW}
\end{eqnarray} 

\begin{figure}
  \includegraphics[width=\columnwidth]{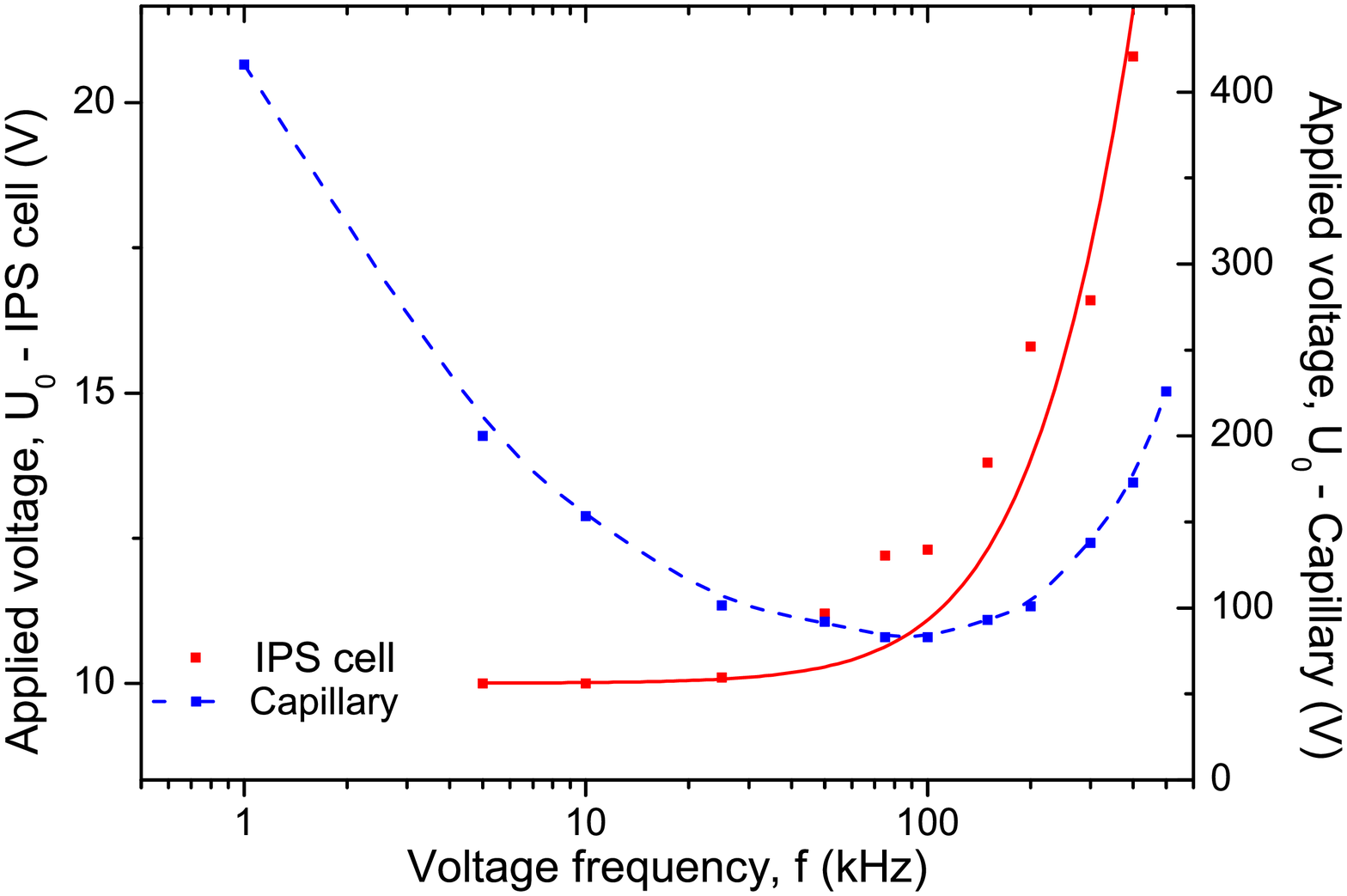}\\
  \caption{\label{Fig:BireCste} Applied voltage as a function of the frequency to maintain constant the Kerr induced birefringence  of $\Delta n = 3.5 \times 10^{-4}$ and of $\Delta n = 1 \times 10^{-5}$ in the geometry of an IPS cell ($h$ = 20~$\mu$m and $d=15~\mu$m ; red symbols) and of a cylindrical capillary ($h=1~$mm and $d=461~\mu$m ; blue symbols), respectively. The sample concentration is $\Phi =$0.66 \%. %(batch 2). 
The red solid line corresponds to the numerical fit using Eq. \ref{eq:fit} and the dashed blue line is a guide for the eye. }
\end{figure}

Figure \ref{Fig:BireCste} displays the operating voltage necessary to apply %in both setups
 in order to maintain constant a given birefringence %($\Delta n = 3.5 \times$ $10^{-4}$)
while the frequency is varied. The main difference between the two setups (Figs. \ref{Fig:IPS} and \ref{Fig:SetupCap}) occurs in the low frequency regime where a strong screening of the field due to the quartz capillary wall is observed. A similar behavior between the two setups is however observed for high frequencies, which is attributed to Debye-type relaxation of the polarization mechanism. Therefore, in the case of  externally applied electric field using quartz capillary, the Kerr coefficient determination has been done at $f=100$~kHz, the optimal frequency for which the coupling between the electric field inside the sample and the induced birefringence is the strongest (See ESI).\\

The Maxwell-Wagner polarizability %allows for the determination of the excess polarizability $\Delta \alpha= \alpha_{\|} - \alpha_{\perp}$ 
is at the origin of coupling with the electric field of the rod-like particles having a volume $V_{p}$ and a conductivity $K_p$, and it can be expressed for the directions parallel ($\|$) or perpendicular ($\perp$) to the rod long axis as follows \cite{Saville, Shen,Dozov2011}:  

\begin{eqnarray}
\alpha_{\|,\perp}=
V_p \varepsilon_{0} \varepsilon_e^*
\frac{(\varepsilon_{p}^* - \varepsilon_{e}^*)}{\varepsilon_{e}^* + (\varepsilon_{p}^* - \varepsilon_{e}^*)L_{\|,\perp}}
\label{eq:alpha}
\end{eqnarray}

 where $\varepsilon_{p}^*=\varepsilon_p + i K_p/(\varepsilon_0 \omega)$ and $\varepsilon_{e}^* = \varepsilon_e + i K_e/(\varepsilon_0 \omega)$ represent the complex dielectric permittivities of the particles and the electrolyte respectively, and $L_{\|,\perp}$ are the depolarization factors. To show its Debye-type frequency dependence, the polarizability can be expressed as: 
\begin{eqnarray}
\alpha_{\|,\perp} =  \alpha_{\|,\perp}^\infty + \frac{\alpha_{\|,\perp}^0 - \alpha_{\|,\perp}^\infty}{1+\omega^{2}\tau_{\|,\perp}^{2}}
\label{eq:alphaDebye}
\end{eqnarray}
with $\alpha_{\|,\perp}^0$ and $\alpha_{\|,\perp}^\infty$ the low and the high frequency electric polarizabilites respectively, $\omega$ the angular frequency of the applied sinusoidal electric field ($\omega = 2 \pi f$) and $\tau_{\|,\perp}$ the Maxwell-Wagner relaxation time given by \cite{Stoylov,Shen}:

\begin{eqnarray}
 \tau_{\|,\perp}=\varepsilon_{0} \frac{ \varepsilon_e+ ( \varepsilon_p - \varepsilon_e)L_{\|,\perp}}{K_e + (K_{p} - K_e)L_{\|,\perp}} 
\label{eq:tauMWO}
\end{eqnarray}

The depolarization factors $L_{\|}$ and $L_{\perp}$ are related through $L_{\perp}=(1-L_{\|})/2$ and have been calculated explicitly in the case of prolate spheroids \cite{Stoylov}. In the needle-like limit ($L\gg D$), the depolarization factors can be strongly simplified and approximated as $L_{\|}\approx 0$ and $L_{\perp}\approx \frac{1}{2}$.
In case of slender rod-like particle when only polarization along the rod is considered, Eq. \ref{eq:tauMWO} is reduced to $\tau_{MW} \approx\tau_{\|} \approx \varepsilon_{0} \varepsilon_e/K_e $ which corresponds in this limit to the charge relaxation time $\tau_c$ according to Eqs. \ref{eq:fc} and \ref{eq:fcMW}. Thus, when rod-like particles are long enough (which means $L/D >  K_p/K_e$), a single frequency, corresponding to the charge relaxation frequency $f_c$, accounts for the frequency behavior of needle-like nanoparticle dispersions.

Assuming such kind of Debye-type frequency dependence and because the polarizability of the particles is proportional to the Kerr coefficient $B$,  Eq. \ref{eq:Kerrparabolic} provides then the frequency dependence of the applied voltage ($U_{0}$): 
% \begin{eqnarray}
% \alpha
% \propto 
% \frac{1}{1+\omega^{2}\tau^{2}}
% \label{eq:pol}
% \end{eqnarray}
\begin{eqnarray}
U_{0}
\propto 
\sqrt{1+(f/f_c)^2}
\label{eq:fit}
\end{eqnarray}

The experimental data (red symbols in Fig. \ref{Fig:BireCste}) obtained in the IPS cell have been numerically fitted (red solid line) according to Eq. \ref{eq:fit} to get the charge relaxation frequency of the system:  
\begin{eqnarray}
{f_{c} \simeq ~210~kHz}
\end{eqnarray}
Replacing $f_c$ in Eq. \ref{eq:fc} provides the Debye screening length in our nanorod suspension, $\kappa^{-1}$ = 9~nm. The ionic strength $I$ can then be  directly deduced from Eq. \ref{eq:Debye}, which gives $I = 0.5$~mM in very good agreement with the value obtained from the phase diagram of LaPO\textsubscript{4} nanorod suspensions reported previously \cite{KimAdv2012}.
According to Eq. \ref{eq:fcInt}, the conductivity of the electrolyte can be also extracted: $K_{e}$ = 4.3 $\times 10^{-4}~$S/m at $I = 0.5$~mM. This value is in very good agreement with the single point measurement of the conductivity performed at a rod volume fraction of $\phi$=0.6\%, and which gives 5.4~$\times 10^{-4}~$S/m. 

\subsection{Determination of the Kerr coefficient and comparison with Maxwell-Wagner-O'Konski theory}

\begin{figure}
  \includegraphics[width=\columnwidth]{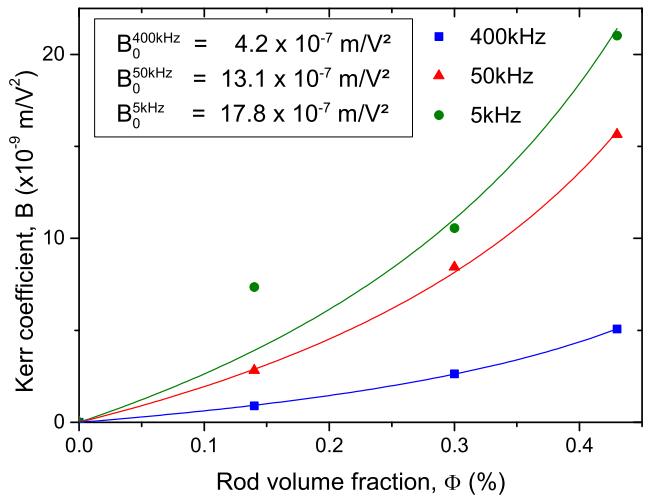}\\
  \caption{\label{Fig:CsteKerr} Dependence of the Kerr coefficient $B$ with the LaPO\textsubscript{4} rod volume fraction obtained at different frequencies. The effective electric field has been used to account for the attenuation due to the IPS cell geometry. The symbols correspond to the experimental data and the solid lines are their corresponding numerical fits (proportional to $\phi/(1-\phi/\phi_B)$) from which the $\phi$-independent specific Kerr coefficient $B_0$ has been extracted. }
\end{figure}

A general expression of the Kerr coefficient has been derived by Saville {\it et al.} \cite{Saville}, and $B$ can be written as a function of the anisotropy of excess polarizability, $\Delta \alpha= \alpha_{\|} - \alpha_{\perp}$:

\begin{eqnarray}
B=
\Delta n^p
\frac{\Delta \alpha}{ 15~\lambda~k_B T }
\frac{\phi}{1-\phi/\phi_B}
\label{eq:Kerr}
\end{eqnarray}
 
 where $\Delta n^{p}$ is the extrapolated saturated birefringence at $\phi$=1, which has been estimated to be $\Delta n^{p} \approx$~0.1 in a previous work \cite{KimAdv2012}. Note that the volume fraction dependence in $\frac{\phi}{1-\phi/\phi_B}$ where $\phi_B$ is the binodal volume fraction at the Isotropic-to-Nematic phase transition, is accounted through an Onsager-like approach \cite{Beek,Shen,Dozov2011}, and leads to a linear dependence with the rod concentration for very dilute suspensions. Fig. \ref{Fig:CsteKerr} displays the evolution of the Kerr coefficient as a function of $\phi$ and plotted for three different frequencies (See ESI for the data of the induced birefringence from which the Kerr coefficients have been obtained). Let's define the  $\phi$-independent specific Kerr coefficient $B_0$ as: 

\begin{eqnarray}
B_0=
\lim_{\phi \to 0}  B(\phi)
\label{eq:B0}
\end{eqnarray}

\begin{figure}
  \includegraphics[width=0.95\columnwidth]{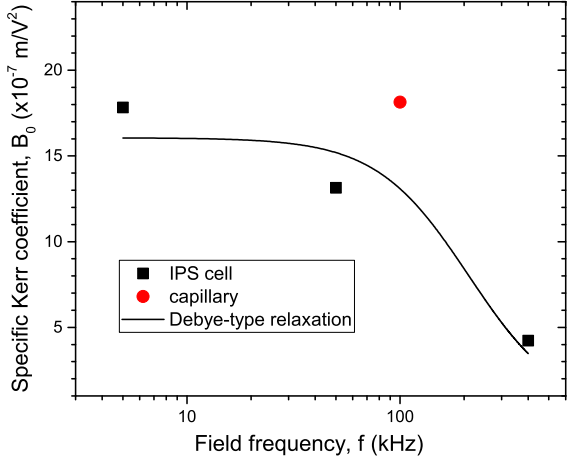}\\
  \caption{\label{Fig:B0} Specific Kerr coefficient $B_0$ at different frequencies obtained in the geometry of directly applied (IPS cell; black squares) or externally applied (quartz capillary; red dot) electric field. In both cases, the effective electric field inside the sample has been calculated to account for its attenuation by the setup. The solid line is a numerical fit corresponding to a Debye-type relaxation (Eq. \ref{eq:alphaDebye}) with $f_c=$~210~kHz. }
\end{figure}

The experimental values of $B_0$ at the three studied frequencies can be found for the IPS cell in the inset of Fig. \ref{Fig:CsteKerr} and in Fig. \ref{Fig:B0}. Similarly, the specific Kerr coefficient has been determined at $f=100$~kHz for externally applied electric field through a capillary wall and is reported in Fig. \ref{Fig:B0}. The latter value is in good agreement with the ones obtained in the IPS cells for the same range of frequency (Fig. \ref{Fig:CsteKerr}) and after correction of the electric field attenuation in both setup geometries.
% Experimentally, $c_g^{\prime 2} \times B_0 =  3.4 \times 10^{-7} m/V^{2}$. We can then extract from the slope of the linear fit (See S.I.) the Kerr coefficient of our LaPO\textsubscript{4} nanorods in the capillary geometry which is  $B_0=18 \times 10^{-7} m/V^{2}$. \\
According to Eq. \ref{eq:Kerr}, comparison between experimental and theoretically predicted Kerr coefficients is done by the determination of the excess polarizability $\Delta \alpha$. %= \alpha_{\|} - \alpha_{\perp}$. 
The Debye-type frequency dependence of the polarizability (Eq. \ref{eq:alphaDebye}) applied for slender rods ($L\gg D$ for which $L_{\|}\approx 0$ and $L_{\perp}\approx \frac{1}{2}$) leads to the following anisotropy of polarizability in the low ($\Delta \alpha^0$) and high ($\Delta \alpha^\infty$) frequency limit, respectively:    

\begin{eqnarray}
\Delta \alpha^0 = V_p \varepsilon_{0}  \frac{\varepsilon_e}{K_e} \frac{(K_p - K_e)^2}{K_p + K_e }
\label{eq:alpha0}
\end{eqnarray}
\begin{eqnarray}
\Delta \alpha^\infty = V_p \varepsilon_{0}  \frac{(\varepsilon_{p} - \varepsilon_{e})^2}{\varepsilon_p + \varepsilon_e}
\label{eq:alphaInf}
\end{eqnarray}

The latter one in Eq. \ref{eq:alphaInf} corresponds to the polarization mechanism due to dielectric contrast between the particles and the surrounding medium. Assuming a typical particle dielectric constant of $\varepsilon_{p} \approx 3$ providing almost the highest dielectric mismatch possible for this system, this yields $\Delta \alpha^\infty /  k_B T\approx 2 \times 10^{-12}~m^2/V^2$ and therefore $B_0^\infty=2 \times 10^{-8}~m/V^2$ according to Eq. \ref{eq:Kerr}, which is more than one order of magnitude smaller than the experimental value at the highest frequency of 400~kHz (Figs. \ref{Fig:CsteKerr} and \ref{Fig:B0}). In a similar way but in the opposite limit of frequency, the contrast of the two conductivities $K_p$ and $K_e$ represents the low frequency limit of the Maxwell-Wagner contribution. The conductivity $K_p$ of the particles has been reported for similar lanthanide phosphate nanorods exhibiting the same morphology: $K_p=6 \times 10^{-4}$~S/m \cite{Wang}. Using $K_{e}$ = 4.3 $\times 10^{-4}~$S/m determined in the previous section, Eq. \ref{eq:alpha0} leads then to $\Delta \alpha^0 /  k_B T\approx 10^{-13}~m^2/V^2$ and therefore $B_0^0=10^{-9}~m/V^2 \ll B_0^{exp}\approx 10^{-6}~m/V^2$ as reported in Figs. \ref{Fig:CsteKerr} and \ref{Fig:B0}. Neither the dielectric contrast nor the contrast of conductivities are able to account for the polarizability excess observed experimentally. However, in the case of charged particles, a third polarization mechanism has to be considered, as first introduced by O'Konski who has shown the importance of the electrical double layer polarization \cite{O'Konski}. This results in a strongly enhanced effective conductivity ($K_p^{eff}$) at the electrolyte/particle interface that has to be substituted in the Maxwell-Wagner formalism (Eqs. \ref{eq:alpha}, \ref{eq:alphaDebye} and \ref{eq:tauMWO}):
\begin{eqnarray}
K_{p,\|,\perp}^{eff}=K_p+2K^{\sigma}/a_{\|,\perp}
\label{eq:Keff}
\end{eqnarray}
 where $K^{\sigma}$ represents the surface contribution to the conductivity and $a_{\|,\perp}$ is an effective length depending on the geometric factors of the particle \cite{Stoylov}. Taking advantage from the high aspect ratio of the LaPO\textsubscript{4} rods ($L\gg D$), only the polarization along the rod long axis %(denoted $\|$)
is considered, neglecting then the normal %($\perp$)
 polarization. In these conditions, the effective conductivity can be expressed as $K_p^{eff}=K_p+2K^{\sigma}/R$ where $a_{\|}\equiv R=D/2$ is the radius of the LaPO\textsubscript{4} particle. Assuming that the surface conductivity can be approximated by \cite{Dozov2011,Shen}:

\begin{eqnarray}
K^{\sigma}\simeq \mu^{EG}_{NO_{3}^{-}}~q^{EG}     
\label{eq:Ksigma}
\end{eqnarray}

with $q^{EG}$ the surface charge density of the LaPO\textsubscript{4}  rod in ethylene glycol related to the zeta-potential $\zeta$ for monovalent electrolyte through \cite{Saville}:

\begin{eqnarray}
q^{EG}=2\frac{\varepsilon_0\varepsilon_e\kappa~k_B T}{e}  
sinh
\left(
\frac{e\zeta}{2 k_B T}
\right)   
\label{eq:qEG}
\end{eqnarray} 

this gives $q^{EG}=6 \times 10^{-3}~C/m^2$ with $\zeta$ = + 97 mV, and therefore a surface conductivity $K^{\sigma}=2.5 \times 10^{-11}$~S. Since $K_p\ll K^{\sigma}/R$, the equivalent conductivity contrast related to the so-called Dukhin number \cite{Stoylov}, gives $K_p^{eff}/K_e \simeq 2/R \times K^{\sigma}/K_e\simeq 21$. According to the extended Maxwell-Wagner-O'Konski (MWO) theory where the conductivity is replaced in Eq. \ref{eq:alpha0} by the effective conductivity $K_p^{eff}$ \cite{Saville, Dozov2011, Stoylov}, the associated excess polarizability can be written as: 
\begin{eqnarray}
\Delta \alpha^{MWO} \approx \alpha_{\|}^{MWO}  \approx V_p \varepsilon_{0}  \varepsilon_e \frac{K_p^{eff}}{K_e} 
\label{eq:alphaMWO}
\end{eqnarray}

This provides $\Delta \alpha^{MWO}/k_B T \approx 3 \times 10^{-11}~m^2/V^2$ and therefore $B_0^{MWO}=4 \times 10^{-7}~m/V^2$ according to Eq. \ref{eq:Kerr}. This value is very consistent with the specific Kerr coefficients reported experimentally in Fig. \ref{Fig:B0}, especially by considering the approximations we performed. Furthermore, this demonstrates that the polarization of the electric double layer is the prevailing mechanism for the coupling with the electric field of our LaPO\textsubscript{4} colloidal rods dispersed in ethylene glycol. \\ 

The corresponding cut-off frequency $f^{MWO}_{\|,\perp}$ is related to the relaxation time $\tau_{\|,\perp}$ (Eq. \ref{eq:alpha}) which has to be written within the extended Maxwell-Wagner-O'Konski model \cite{Stoylov,Shen}. In case of slender rod-like particles when only polarization along the rods is considered and which satisfy $L/D >  K_p^{eff}/K_e \gg 1$, Eq. \ref{eq:tauMWO} is simplified into $f^{MWO}\approx f^{MWO}_{\|}\approx f^{MW}_{\|}\approx K_e / (2\pi \varepsilon_{0} \varepsilon_e) \equiv f_c$, which corresponds again to the charge relaxation frequency $f_c$ reported in Eq. \ref{eq:fcMW}. Therefore a numerical fit assuming a Debye-type relaxation (Eq. \ref{eq:alphaDebye}) with  $f_c=$~210~kHz has been performed as reported in Fig. \ref{Fig:B0}, showing that polarization of the particle ionic cloud does not only account for the right amplitude of the Kerr effect coefficients, but also accounts very well for their frequency dependence. 

%This is consistent with the change of polarization mechanism discussed above, from a prevailing polarization of the particle ionic cloud at lower frequencies to an interfacial electric polarizablility at higher frequencies. 
%It has to be noted that the dilute volume fraction of the suspensions used in our study allows us to neglect the phenomenon related to the electro-osmotic flow.

\subsection{Switching time (On/Off states)}

\begin{figure}
  \includegraphics[width=\columnwidth]{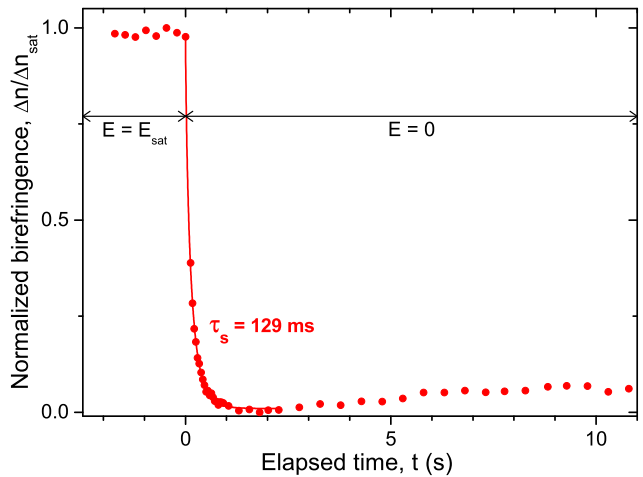}\\
  \caption{\label{Fig:Rlx} Switching time %or time response 
%of the normalized light intensity ($I / I_0$) associated to 
associated with the relaxation of the Kerr induced birefringence when the electric field is turned off. Experiment performed in an IPS cell at $f=5$~kHz for a sample having a rod volume fraction of $\phi$=0.66\%. The red solid line represents the numerical fit using a stretched-exponential decay (Eq. \ref{eq:ExpDecay}) and the corresponding time scale is provided. %zero time marks the start of the relaxation process when the electric filed is switched off.
}
\end{figure}

Beyond the coupling with the electric field, investigations on the time response of the LaPO\textsubscript{4} colloidal rod suspensions have been performed, as displayed in Fig. \ref{Fig:Rlx}. For a frequency of 5 kHz, the induced birefringence is brought to saturation ($\Delta n_{sat}$, see Fig. \ref{Fig:Freq}) in a steady-state regime. The applied voltage is then quickly switched off (setting the origin of time in Fig. \ref{Fig:Rlx}) and the relaxation into the isotropic liquid state (i.e. with no birefringence) occurs in two steps. First, a fast decay of the transmitted light intensity takes place when the particles lose their orientation; second, a slight increase of the birefringence is observed, as already reported in other self-organized systems \cite{Eremin}. We interpret the latter behavior as a reminiscent planar anchoring of the particles on the glass substrates, which leads then to a residual nematic-like order close to the interface.    

From the numerical fit using a stretched-exponential form of the type \cite{Degiorgio}: 
\begin{eqnarray}
 \Delta n=\Delta n_{sat}exp(-(t/\tau_s)^\gamma),~~~ 0\leq \gamma \leq 1
\label{eq:ExpDecay}
\end{eqnarray}

the time response of the suspensions has been extracted %at two frequencies, 5 and 50 kHz for which $\Delta n_{sat}$ can be obtained, 
giving $\tau_s=$129~ms (with $\gamma \simeq 0.8$) as shown in Fig. \ref{Fig:Rlx}. The rod polydispersity leads to a distribution of relaxation times implying the expression used in Eq. \ref{eq:ExpDecay}, reducing to $\gamma = 1$ in the simplest case of monodisperse particles.  
The switching time $\tau_s$ found in our system is too slow for a use of these colloidal suspensions in displays but is fast enough to consider potential application in devices such as smart windows or lenses. Note that the relaxation of the Kerr birefringence back to the isotropic liquid state stems from the thermal diffusion randomizing the particle orientation initially driven by the electric field. Therefore, the rotational diffusion coefficient ($D_r$) can be extracted from the time response $\tau_s$ via  $\tau_s=1/6D_r$ \cite{Jimenez,Peikov}. 
 As the rod volume fraction used in the experiment ($\phi$=0.66\%) is close to the binodal value and corresponds to the semi-dilute regime where the motion of each rod is hindered by its neighbors, this results in a decrease of the rotational diffusion compared to free rods, and $D_r$ has then the following form \cite{Pecora}:
 
 \begin{eqnarray}
 D_r=\beta k_B T \frac {ln(L/d)}{\eta_{EG} L^9 (\phi/v_p)^2}
\label{eq:Dr}
\end{eqnarray}
 
where $\beta$ is a proportionality constant. Eq. \ref{eq:Dr} yields an associated time of about 100~ms with $\beta \simeq 0.5$ consistent with the response time $\tau_s$ obtained experimentally (Fig. \ref{Fig:Rlx}).  %In order to get a better agreement, the size polydispersity of the LaPO\textsubscript{4} rods which has been neglected here should certainly be included in the calculation of the rotational diffusion coefficient \cite{Peikov}.    

\section{Conclusions}

% \textbf{Penser à mentionner l'effet d'une petite fraction d'H20 dans nos suspensions d'EG qui pourrait se concentrer autour de nos NPs et influencer l'effet d'orientation de notre EDL}

In this paper, we present a study of the Kerr induced birefringence in concentrated isotropic liquid suspensions of LaPO\textsubscript{4} nanorods in ethylene glycol. The response of the system was probed both in an In-Plane Switching cells in which the electrodes are in direct contact with the sample, and also in the geometry of externally applied field. In the latter case, the electric field is applied through and parallel to a capillary quartz wall, limiting its attenuation and avoiding electric damage such as charge injection at the electrodes. %electrolysis due to the electrolyte mobile charges reaching the electrodes. 
A quantitative analysis using the Maxwell-Wagner-O'Konski theory shows that the Kerr birefringence is mainly the result of the polarization of the electric double layer of the rod-like particles. The typical value of the specific (i.e. $\phi \to 0$) Kerr coefficient obtained in our system is of about $B_0\sim 10^{-6} m/V^2$ which is several orders of magnitude higher than usual thermotropic compounds \cite{Ghana} including liquid-crystalline blue phases \cite{YanAPL2010,Hisakado,Nordendorf}.  Moreover, the coupling with the electric field is also very similar to other aqueous systems of rod-like or plate-like particles \cite{Bellini99,Dozov2011,Jimenez} showing therefore that ethylene glycol as electrolyte combined with well-defined anisometric nanoparticles represent a promising way for their use in electro-optical devices.\\
%The frequency dependence study highlights. We then focused on the volume fraction dependence of the coupling between the electric birefringence and field as well as the time response of the dispersions when relaxing from an induced nematic to the isotropic liquid state. The time constants obtained (of about 100 ms) are fast enough to consider potential applications in electro-optical devices such as smart windows or lenses.\\

\begin{acknowledgments}
We would like to thank J. Giermanska for her help in the zeta-potential measurement, Essilor Int. for preliminary discussion, and I. Dozov for fruitful comments.  
\end{acknowledgments}

Author contributions: EG designed research; AC performed the experiments and collected the data; PM and EG designed the experimental setups; PM performed the numerical simulations; the LPMC group (JWK, KL, JPB, TG) synthesized the samples; EG and AC interpreted the experiments and wrote the paper. All the authors discussed the manuscript.

\end{document}